\documentstyle[11pt,paspconf,epsf]{article}
\def \etal         {{\it et~al.}~}

\begin{document}

\title{Dynamical models of S0 and Sa galaxies}
\author{E. Pignatelli and G. Galletta}
\affil{Dipartimento di Astronomia, Universit\`a di Padova, Vicolo
dell'Osservatorio 5, Padova, Italy}

\begin{abstract}
We present a set of detailed, self-consistent, isotropic dynamical models for
disc galaxies.  We start from the hypothesis that each galaxy can be decomposed
in a bulge, following the $r^{1/4}$ law, and a disc with an exponential
projected density profile; and that the isodensity surfaces of each component
can be represented by similar concentric spheroids.  Under these conditions we
produce the rotational velocity and velocity dispersion profiles, after taking
into account both the asymmetric drift effects and the integration along the
line of sight.

The model is successfully applied to reproduce the stellar kinematic
and photometry of the bulge of the 4 lenticular and early-type spiral
galaxies NGC~4565, NGC~7814, NGC~5866 and NGC~4594.  For these
galaxies detailed stellar kinematical data are available at different
positions across the galaxy. 

The application of our models shows that: 1) For all the galaxies
considered in this work, an isotropic model is able to reproduce the
whole dynamical data.  This is surprising in the special case of
NGC~4565, for which previous models were unable to reproduce the
velocity distribution without introducing strong anisotropies.  2) We
do not need a dark mass halo to reproduce these data.  This do not
mean that a dark halo is not present, but just that its dynamical
effects in these inner regions of the galaxies are negligible.

In our opinion, these results strongly support our guess that a complete 
dynamical model, including {\em both} a disc and a bulge component, is needed 
to understand the structure of the lenticular and Sa galaxies. 
\end{abstract}

\section{Introduction}

The study of the distribution function of stars in galaxies may give strong
constraints on the scenarios of formation of the galaxies.  In the past, a
particular attention has been paid to distinguish between isotropic cases,
where the distribution function is characterized by only two classical
integrals of motion (energy and angular momentum) and the anisotropic case,
where a third (unknown) isolating integral is needed to explain the dynamics
of the system.

A widely used tool to investigate the distribution function in elliptical
galaxies is the classical $V/\sigma$ parameter, first introduced by Binney
(1978). By applying this method to a large number of elliptical
galaxies, it seems that two, different kind of ellipticals may exist:
elliptical galaxies fainter than M$_B$=19.5, with outer isophotes quite flat
(disky), which are represented well by isotropic models, and bigger
ellipticals (M$_B \le$ 19.5), with isophotes more ``boxy'' and that cannot
be explained by a model with isotropic velocity ellipsoid.

The structure and isotropy of the bulges of disk galaxies is far less clear:
working on the same set of data, concerning fours S0 and Sa galaxies,
Kormendy \& Illingworth (1982) claims the isotropy of their bulges,
while Whitmore, Rubin and Ford (1985) deduce the opposite conclusion.
The fact that the same data may be interpreted in such different ways
clearly demonstrates the ambiguity of the $V/\sigma$ test when applied to
the bulges.

On the other hand, the informations about the dynamical structures of the
bulges could be crucial to determine if bulges can be considered as ``dwarf
ellipticals embedded in a gas disk'', as stellar populations, colors and
luminosity profiles seems to suggest, or if they are a totally different
class of objects.

\section{The model}

\begin{figure}[t]
\plottwo{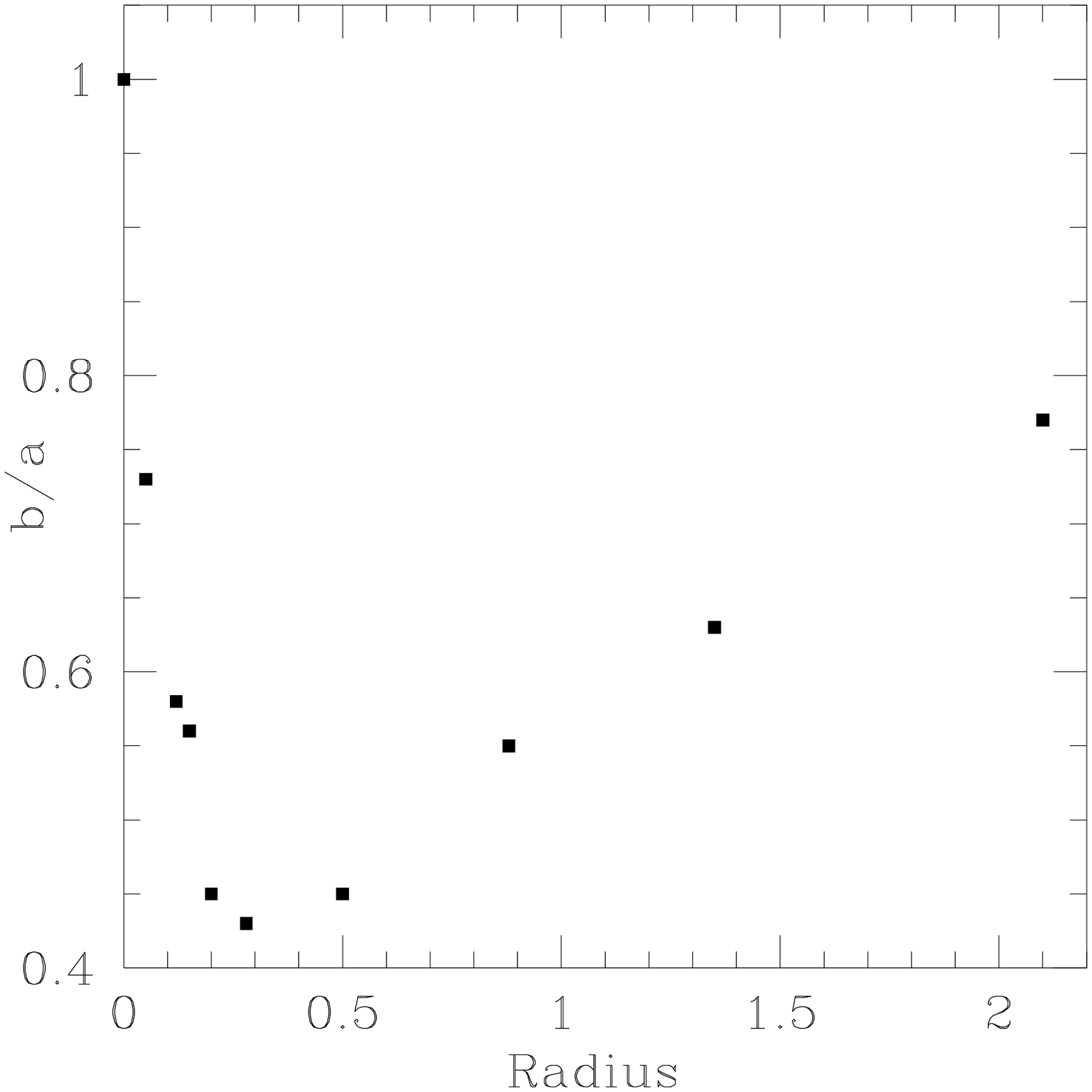}{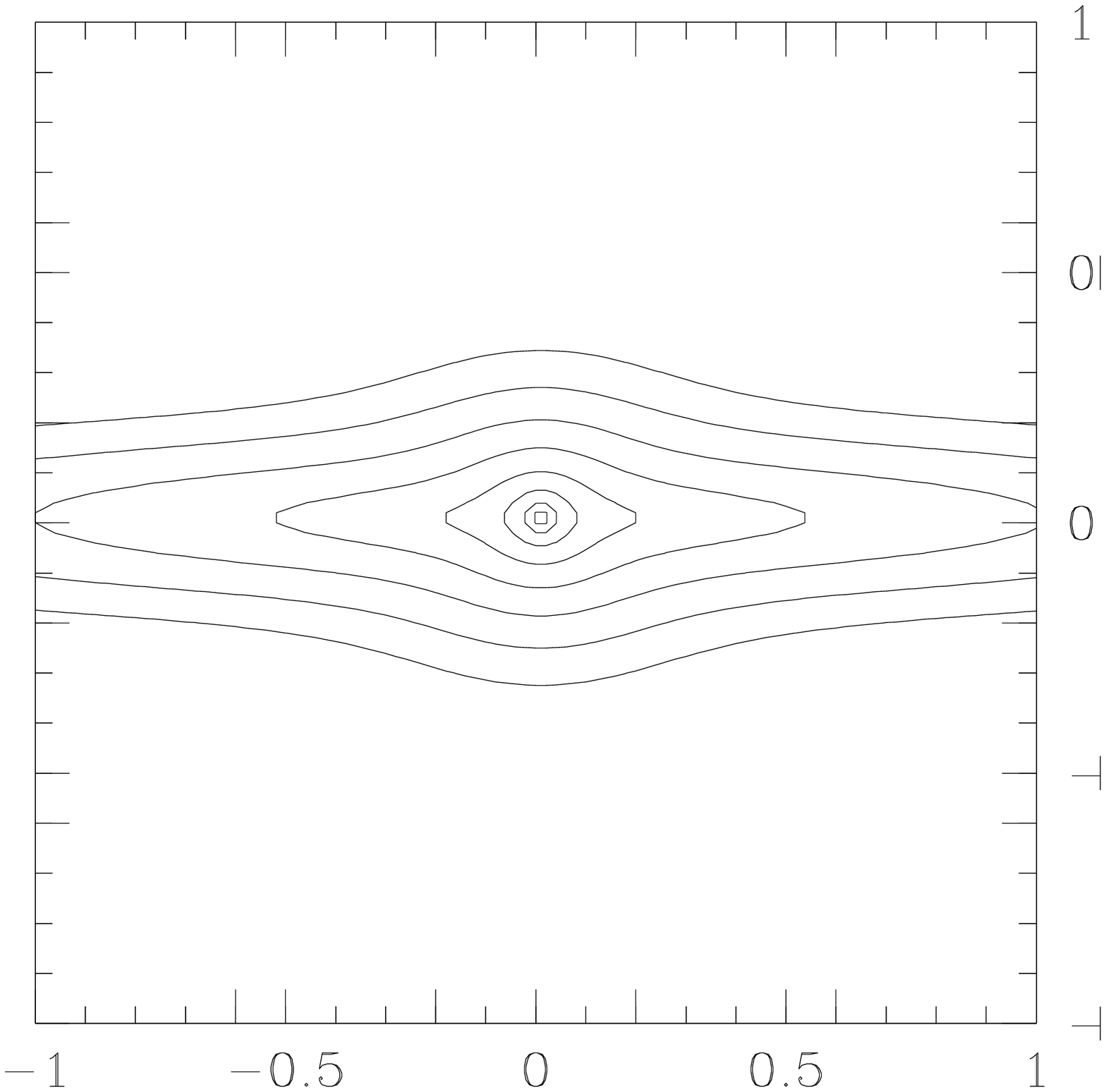}
\caption{ Photometry of a model with two components having different axial
ratios. We assume a spherical bulge and an exponential disk with 
 $b/a = 0.1$.  On the left figure we show the isophotes of such a model,
while on the right is shown the behavior of the global axial ratio as
a function of the radius. Notice how the isophotes becomes more and
more flatter while moving out from the center, and then turn back to
be round when the bulge component become dominant again. All radii are
in units of $R_e$.}
\label{fig:iso}
\end{figure}

The main problems encountered dealing with models of bulges is the mixing of
stars between disk and bulge and the mutual interaction of their potentials.
These effects make harder the problem of to study the ``pure'' bulge
component, that require some preliminary assumptions.

\newpage

\begin{figure}[h]
\plotone{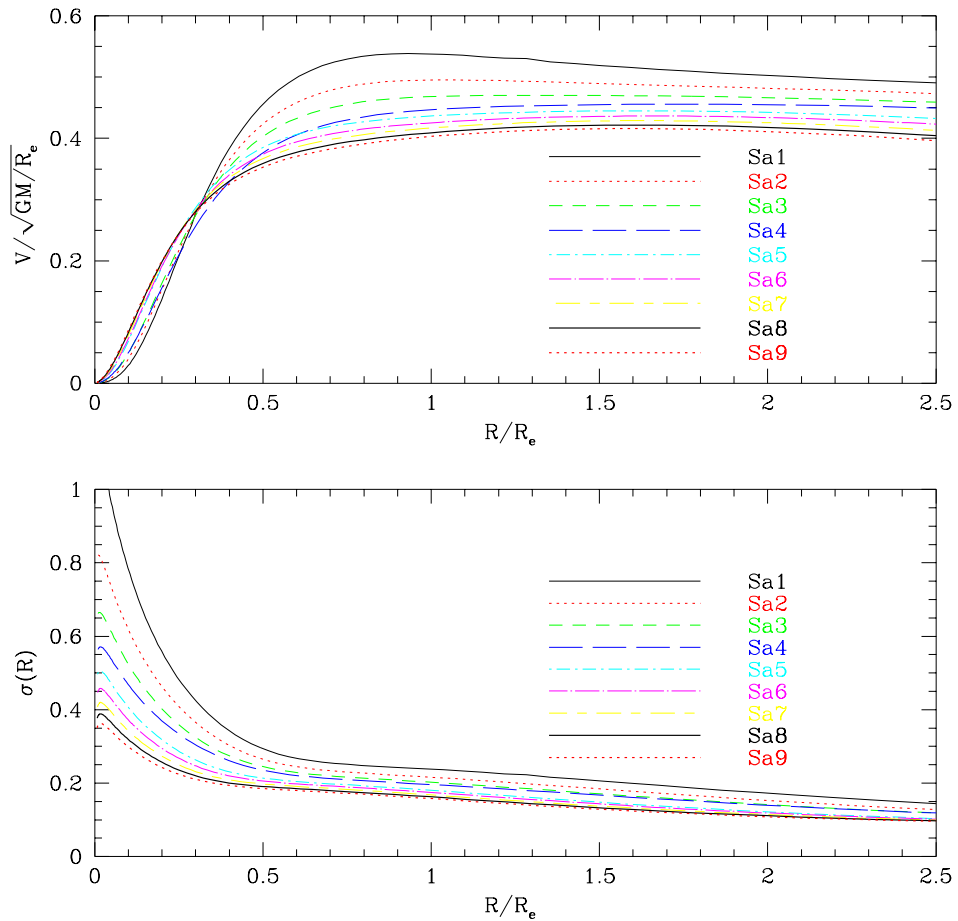}
\caption{Rotational curves and velocity dispersion profiles for models with
different values of the parameters $R_B/R_D$, $M_B/M_D$. Here we adopted the
values $M_B/M_D=1.36$ which, according to Simien and de Vaucouleurs (1986),
corresponds to a typical Sa galaxy, while the $R_B/R_D$ parameter is ranging
from 0.1 to 0.9 (models Sa1 through Sa9). }
\label{fig:sa}
\end{figure}

In modeling the elliptical galaxies it is common to assume that the
isodensity surfaces are similar, concentric spheroids. This is in reasonable
agreement with the shape of the observed isophotes, and has the great
advantage to reduce the complexity of the dynamical equations. As said before,
in such a case (ellipsoidal structures) the $V/\sigma$ test can be applied.

On the other hand, the isophotes of bulges in disk galaxies shows many
different behaviors, such as wide change of flattening from the center
of the galaxy to the outer regions. In this case, it is not possible to apply
a simple ellipsoidal model. This problem is particularly presenting S0 and Sa
galaxies, where the contributions of the bulge- and disk- components are
comparable in strength.

\newpage

Starting from a previous work of Galletta \etal (1990), we extended the
ellipsoidal hypothesis to early-type disk galaxies, assuming that each galaxy
may be ``splitted'' in two different components, having each one the following
properties:

\begin{itemize}
\item The isodensity surfaces are similar concentric spheroids;
\item The anisotropy parameter $\beta$ is constant throughout the
galaxy (Here, we present only the case $\beta=0$ of the isotropic model).
\item The galaxy is rotating about the $z$-axis with velocity $V(R,Z)$, 
consistent with the self-gravity hypothesis.  
\item Each component has a constant $M/L$ ratio.
\end{itemize}

Under these hypothesis, the circular
velocity due to the potential of each component is given by the well-known form
(see Binney \& Tremaine, 1986):
\begin{equation}
\label{eq:vcell}
v_c^2(R,z)= - 2 \pi G \sqrt{1-e^2} R\int_0^\infty 
         \frac{\rho(m^2) \,d\tau}{(\tau +1)^2 \sqrt{\tau+1-e^2}} 
\end{equation}
with $e$ the ellipticity of the component, and $m$ defined as
\begin{equation}
\label{eq:ellm}
m^2=\frac{R^2}{\tau+1} +\frac{z^2}{\tau+1-e^2}
\end{equation}

The density profile of the bulge is assumed to follow the $r^{1/4}$ law, while
the disc (also ``ellipsoidal'', but much flatter than the bulge) is described
with an exponential density profile.  In Fig.~\ref{fig:iso} we present the
photometric behavior of a typical galaxy constructed with these assumptions.

{\em Since we are dealing with the inner kinematical properties of the
galaxies, we deliberately discard any dark matter contribution to the overall
potential, to limit the number of free parameters. }

\section{A library of models}

\begin{figure}[h]
{\centering \leavevmode \epsfxsize=.9\textwidth \epsfbox{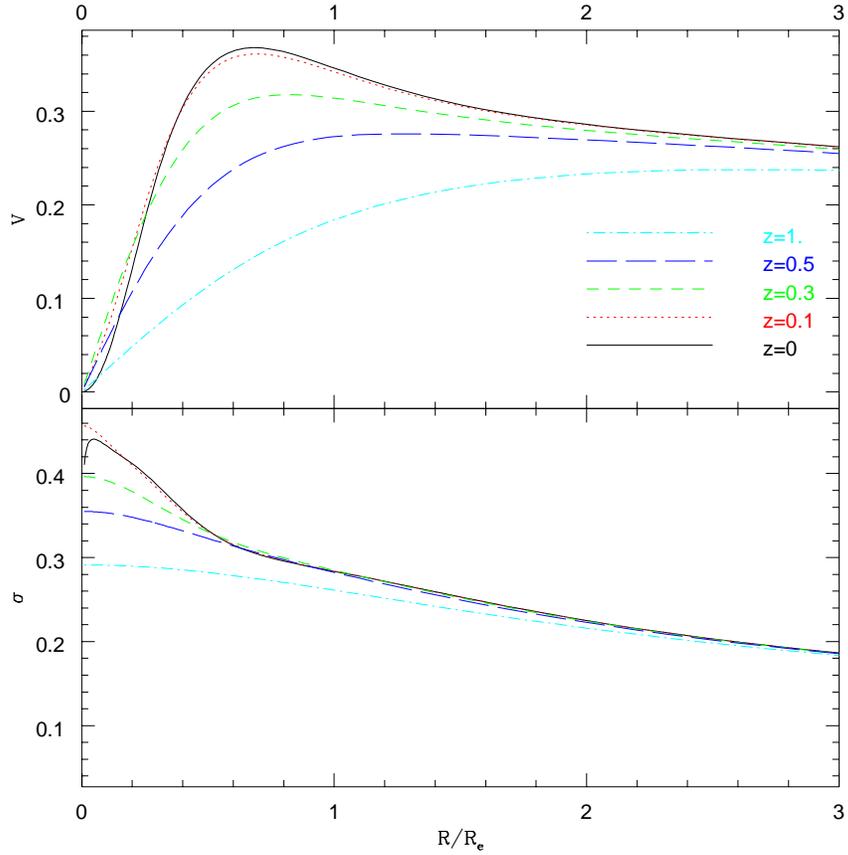} }
\caption{Behavior of the rotational velocity and velocity dispersion profile in
a pure bulge model with a given axial ratio $b/a=0.6$. The different profiles
are extracted at different vertical shifts $z$ from the major axis, as shown in
the legend. Both the asymmetric drift effect and the line-of-sight projection
are taken into account. All the models are seen edge-on and along the major
axis.}
\label{fig:bulge_h}
\end{figure}

A first goal of this work was to find out how kinematical (rotation
and velocity dispersion) curves are changing when different bulge/disk
ratios are assumed.  Under the hypotheses described in the previous
section, we produced several different models for different values of
the parameters.  In Fig.~\ref{fig:sa} and in Fig.~\ref{fig:bulge_h}
some of the obtained results are shown.

We want to stress out some points:

\begin{itemize}

\item The rotation curves of the galaxies on a plane perpendicular to the
main galaxy plane become more and more linear at higher distance from this
latter plane. This is in agreement with the available data for most of the
observed spiral bulges;

\item On the other hand, the asymmetric drift effect is very strong in the 
nucleus; this is due to the steep increase of density for the inner regions
of the bulge (described as a pure $r^{1/4}$ model, without any core radius).

\item The velocity dispersion profile is always {\bf not} gaussian;
      more likely, it can be approximated by the sum of two gaussian curves.
\end{itemize}
 
\section{ Comparison to the observations }

\begin{figure}[h]
{\centering \leavevmode \epsfxsize=.95\textwidth \epsfbox{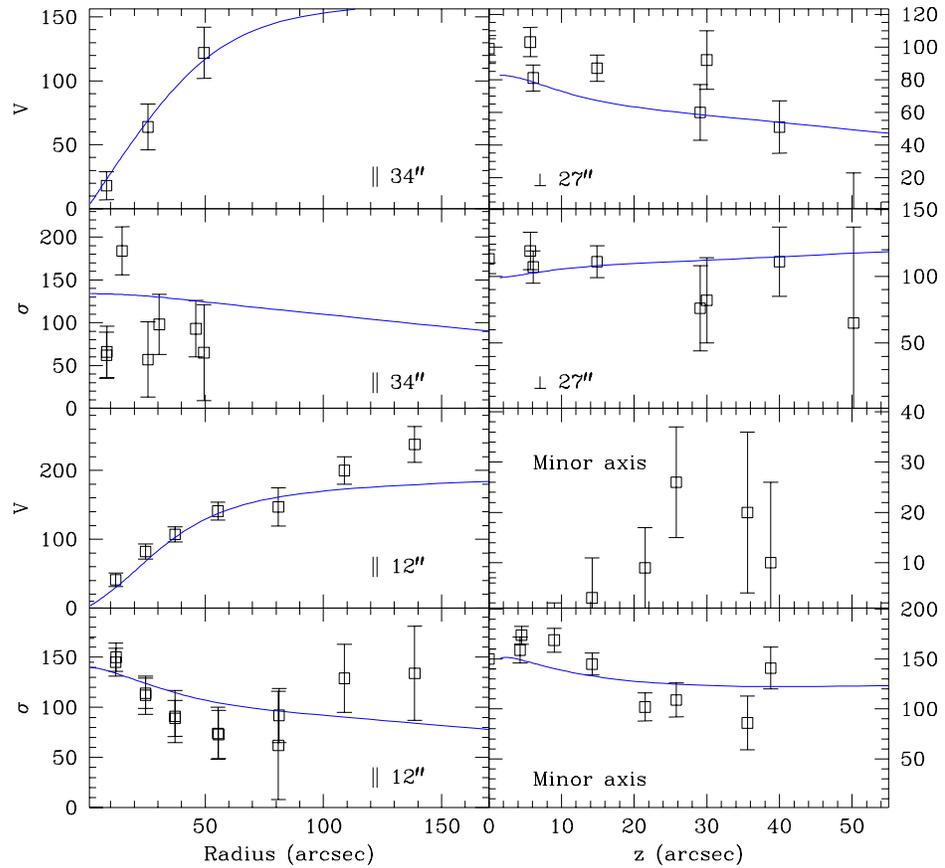} }
\caption{ The rotational curves and velocity dispersion profiles predicted by
the model compared with the observations of Kormendy \& Illingworth
(1982).  The label $\parallel$ means that the data are extracted along
an axis which is parallel to the major axis, while the label $\perp$
label means parallel to the minor axis. All the velocities and
velocity dispersions are in Km/sec.  As described into the text, the
model is described by {\em one} free parameter only.}
\label{fig:n4565}
\end{figure}

As a second goal, we applied our model to the galaxies studied by
Kormendy and Illingworth (1982) and by Whitmore \etal (1985).  In
Tab.~1 the best-fit parameters found for all the four galaxies are
shown.  Our isotropic model fit fairly well the observations; this is
particularly amazing in the case of NGC~4565 (Fig.~\ref{fig:n4565}),
for which both the papers conclude that strong anisotropies in the
velocity distribution are present.

\newpage 

\begin{table}[t]
\caption{Best-fit parameters of the model for the four early-type disk galaxies
studied by Kormendy \& Illingworth.} 
\label{KItab}
\begin{center}
\begin{tabular}{c|cccc}
\tableline
& NGC 4565 & NGC 4594 & NGC 5866 & NGC 7814 \\
\tableline
Hubble Type   & Sb    & Sa      & S0 & Sab \\
$L_B/L_D$& 0.538 & 11.5    &  9 & 1.63 \\     
$R_B$ & $50.59''$ & $18.9''$ & $32.2''$ & $31.3''$ \\
$R_D$ & $122.5''$& $90''  $ & $41''   $ & $117''  $ \\
$(b/a)_{B1}$ &0.59 &0.55 &0.47 & 0.53 \\
 D(Mpc) &  18.9 & 19.3 & 18.4 & 25.0 \\
$M_B$ &  -21.11 & -22.26 &  -20.47 & -20.73 \\
$(M/L)_B$ & 9.5 &  7.5 &  7. & 12. \\
$M_{tot}$ & 4.17 &8.6 &1.72 &4.04 \\
$(b/a)_{B}$ & 0.59 & 0.55 & 0.8 & 0.55 \\
$R_B/R_D$ & 0.7 & 1.0 & 1.0 & 0.487 \\
$M_B/M_D$ & 1.06 & 1.0 & 9.0 & 1.63 \\
\tableline
\tableline
\end{tabular}
\end{center}
\end{table}

The advantage of our model is that the only free parameter left is the
$M/L$ ratio of the stellar component, if a detailed photometry is available.
In such a case, the decomposition of the galaxy into a bulge and a disk
component can be properly made, and we are able to predict the observed
kinematics and photometry at any point of the galaxy on the sky simply by
changing the value of this parameter. The diagnostic power of the model is,
under these conditions, very high.

\section{Conclusions}

We developed a dynamical model to study the presence of isotropy in the
bulges of early-type disk galaxies. The main advantage on previous works is
that both the bulge and disk contributions may be properly taken into account,
as well as the presence of the asymmetric drift and the integration along the
line-of-sight.

The application of our model shows that:
\begin{itemize}
\item For all the galaxies considered in this work, an isotropic model is able
to reproduce the whole dynamical data. This is surprising in the
special case of NGC~4565, for which previous works suggested strong
anisotropies in the velocity distribution.
\item The model don't needs a dark halo to reproduce these data. Its presence
is, of course, possible, but its dynamical effects are negligible in the inner
regions ( $R\le 1.2 R_e$ ) of the considered galaxies. This result is in
agreement with the conclusions of other authors (e.g. Persic \& Salucci, 1995).
\item The method seems to have a great diagnostic power if good kinematical 
data at different position angles are available and/or if many offset
spectra have been taken.
\end{itemize}

\acknowledgments
This work has been partially supported by the grant `Astrofisica e
Fisica Cosmica' Fondi 40\% of the Italian Ministry of University and
Scientific and Technologic Research (MURST).

\end{document}